\documentclass[aps,prl,showpacs,amsfonts,amssymb,amsmath,floatfix,twocolumn,
floats,nobalancelastpage]{revtex4}

\usepackage{graphicx}

\begin{document}          

\title{Universal Scaling in Mixing Correlated Growth with Randomness}

\author{A. Kolakowska}
\author{M. A. Novotny}
\author{P. S. Verma}
\affiliation{Department of Physics and Astronomy, and the ERC Center for 
Computational Sciences,
P.O. Box 5167, Mississippi State, MS 39762-5167}

\date{\today}

\begin{abstract}
We study two-component growth that mixes random deposition (RD) with a 
correlated growth process that occurs with probability $p$. We find that 
these composite systems are in the universality class of the correlated growth process.  
For RD blends with either Edwards-Wilkinson or Kardar-Parisi-Zhang processes, 
we identify a nonuniversal parameter in the universal scaling in $p$. 
\end{abstract}


\maketitle

Many properties of complex systems can be uncovered by statistical 
analysis of some representative nonequilibrium interfaces. Mainstream 
studies of surface growth and interface roughening focus on one-component 
growth, or homoepitaxy, and large-scale properties. On the microscopic level, 
nonequilibrium interfaces have been studied in a variety of discrete 
simulation models such as ballistic deposition (BD), Eden or solid-on-solid 
models. While one-component growths are well understood \cite{BS95}, 
the same can not be claimed about composite systems, even as simple as 
binary growth in one spatial dimension. Several mixed-growth models  
studied in the recent decade \cite{HA03,mix10} reveal new and 
nontrivial properties. But the theory behind these is in the initial stages.

In this first systematic study of two-component growth, we examine a system 
whose dynamics is governed by two simultaneously present processes: 
one is a process that builds up correlations  
(a pure-correlated growth) and the other process is totally 
uncorrelated, i.e., random deposition (RD). The pure-correlated growth 
occurs with probability $p$. Questions that we address 
here concern the universality of such composite systems. As we shall show, 
the presence of randomness slows down the dynamics of the correlation  
processes. Nevertheless, the universality class of the combined 
processes is the same as the universality class of a correlation process. 
This is an outcome of scaling in $p$.  
One consequence of this observation 
is a magnifying-glass effect that RD-blending has on the time-evolution 
of the surface roughness. 
This effect can be useful in revealing hidden features of a correlated growth 
when designing simulation models. Intuitively, since RD carries no correlations 
of its own, it may be expected that its admixtures should not lead to a 
new universality class. Yet, demonstration of this is not so trivial 
since, as we shall make evident by the results of several simulations, 
some of the parameters involved in the universal scaling may be nonuniversal. 
Results presented here for $(1+1)$ dimensions can be easily extended to 
multidimensions.

Consider aggregation models where particles fall onto a one-dimensional 
substrate of $L$ sites, where they may be accepted in accordance to a  
rule that generates correlations among the sites. This pure-correlated 
growth occurs with probability $p$ and competes with RD growth that occurs with 
probability $q=1-p$. When a particle is accepted at a site, 
the site increases its height by $\Delta h$. 
Roughness of the growing surface is measured by the interface width $w(t)$ 
at time $t$: 
$\langle w^2(t) \rangle = \langle L^{-1} \sum_{k=1}^{k=L} [h_k(t)-\bar{h}(t)]^2 \rangle$, 
where $h_k(t)$ is the height at site $k$ and $\bar{h}(t)$ is 
its mean over $L$ sites (angular brackets denote the mean 
over $N$ configurations).

In a pure-correlated growth ($p=1$), assuming elementary linear and 
nonlinear models, the self-affined roughness obeys 
the Family-Vicsek (FV) scaling \cite{FV85},
\begin{equation}
\label{FV-1}
w^2(t) = L^{2\alpha} F(t/L^z),
\end{equation}
where $F(y)$ gives two evolution limits:  
$F(y) \sim y^{2\alpha/z}$ if $y \ll 1$ (growth); and, $F(y) \sim \textrm{const}$ if 
$y \gg 1$ (saturation). The cross-over time $t_\times$ from growth 
to saturation is given by the dynamic exponent 
$z$, $t_\times \sim L^z$ (Fig.~1). At saturation the width 
does not depend on time, $w^2 \sim L^{2\alpha}$, where $\alpha$ is the 
roughness exponent. During the growth $w^2(t) \sim t^{2\beta}$, where  
$\beta = \alpha/z$. Exponents $z$, $\alpha$ and $\beta$ are universal. 
This means, two different simulation models 
of two different correlation mechanisms will generate the same type 
of scaling, with consistent values of exponents, provided these mechanisms 
represent the same type of correlation process, i.e., belong to one universality 
class. 
Dynamics of the buildup of correlations and dynamical scaling are described 
within a continuum model by a stochastic growth equation. One example  
is the Kardar-Parisi-Zhang (KPZ) equation \cite{KPZ86}
\begin{equation}
\label{KPZ-1}
h_t = v(t) + \nu_0 h_{xx} + ( \lambda_0 /2 ) h_{x}^2 + \eta (x,t),
\end{equation}
where $h=h(x,t)$ is the height field (subscripts denote partial derivatives; 
$x$ is the coordinate along the substrate), $v$ is the mean interface 
velocity, and $\eta$ is the white noise ($\nu_0$ and $\lambda_0$ are coefficients). 
In the KPZ universality class, governed by Eq.~(\ref{KPZ-1}),   
$\alpha + z = 2$ and $\alpha = 1/2$. 
When $\lambda_0=0$, Eq.~(\ref{KPZ-1}) becomes the Edwards-Wilkinson (EW) equation \cite{EW82}, 
defining the EW universality with $2\alpha +1 =z$ and  
$\alpha = 1/2$. When $\nu_0=\lambda_0=0$, 
Eq.~(\ref{KPZ-1}) describes uncorrelated processes of RD universality, 
characterized by $\beta =1/2$, $t_\times = \infty$, and the absence of scaling in $L$. 
For EW processes, Eq.~(\ref{FV-1}) expresses the invariance 
of the EW equation under the scaling \cite{BS95}
\begin{equation}
\label{transform-1}
x \to Lx \, , \; h \to L^\alpha h \, , \; t \to L^z t \, .
\end{equation}
Similarly, for KPZ processes it expresses the invariance of the convective 
derivative in the Burger's equation.

%%%%%%%%  figure 1 %%%%%%%
\begin{figure}[tp]
\includegraphics[width=6.0cm]{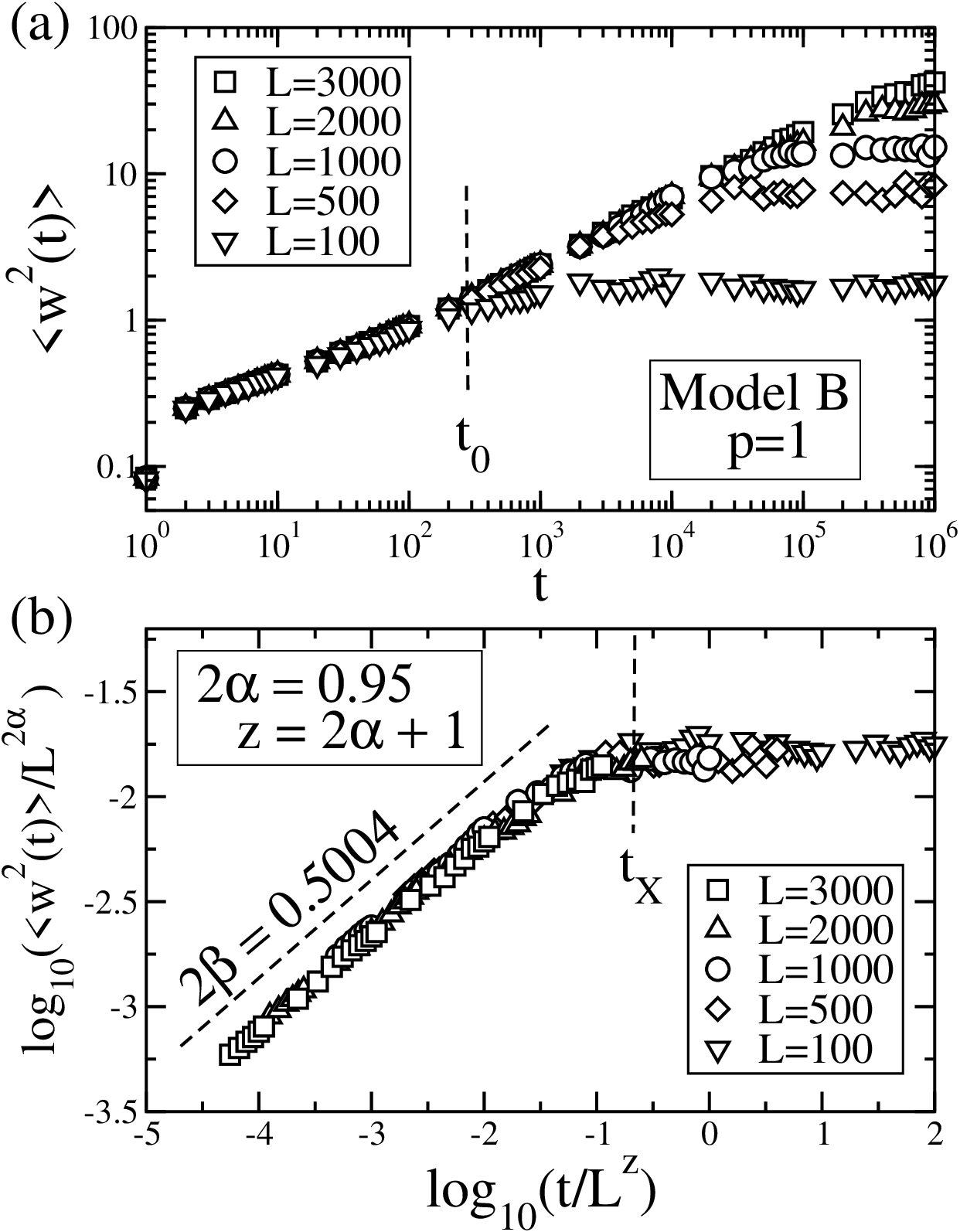}
\caption{\label{fig-1}
{\it Model B} for pure-correlated growth:
(a) Time evolutions of the interface width
($t_0$ marks the end of the initial nonscaling regime);
(b) Scaling function for $t>t_0$.
$z$, $\alpha$ and $\beta$ are consistent with the EW universality.
Here, $N \approx 100$.
}
\end{figure}
%%%%%%%%%%%%%%%%%%%%

In simulations, $t$ is the number 
of deposited monolayers. The first step is RD 
to a flat substrate. The system retains the memory of this 
initial condition for $t_0$ steps, where $t_0$ depends on 
the particulars of the model, i.e., $t_0$ is a nonuniversal 
parameter. In this start-up regime $w(t)$  
does not scale \cite{KNV04}; scaling occurs only for $t>t_0$ (Fig.~1).

In deriving the scaling hypotheses, we are guided by the following four models. 
\textbf{\textit{Model A}}: for $p=1$ is RD with surface relaxation where $\Delta h=1$ 
\cite{BS95,Fam86}, known to be in the EW 
universality class (for $p<1$, studied in \cite{HA03,HMA01}). 
In \textbf{\textit{Model B}}, 
being introduced here, $\Delta h$ is sampled from a uniform distribution of unit 
mean and the substrate is sampled sequentially at each $t$. 
When $p=1$: particles that fall on the local interface minima are 
always accepted; particles that fall on local maxima slide down to either of 
the neighboring sites with probability $1/2$; and, particles that fall on local 
slopes slide down to nearest-neighbor sites. {\it Model B} for $p=1$ 
is in the EW class (Fig.~1). It simulates, e.g., 
deposition of a sticky non-granular material of variable droplet size. 
\textbf{\textit{Model C}}: 
for $p=1$ is BD with $\Delta h=1$, known to be  
of KPZ universality \cite{BS95} (for $p<1$, studied in \cite{HA03,HA01}). 
In \textbf{\textit{Model D}}, $\Delta h$ is sampled from a Poisson distribution of unit mean, 
and each monolayer is obtained by sequential sampling. When $p=1$ 
in {\it Model D}, particles are deposited only to local surface minima. 
This case is in the KPZ universality 
class \cite{KNV04}. {\it Model D} simulates, e.g., conservative 
updates in a system of asynchronous processors  
\cite{KNV04,KN05}. We stress that, although in one universality 
class for $p=1$, {\it Models C} and {\it D} are {\it essentially different} 
simulations (as is the pair {\it A} and {\it B}).

%%%%%%%%  figure 2 %%%%%%%
\begin{figure}[tp]
\includegraphics[width=8.0cm]{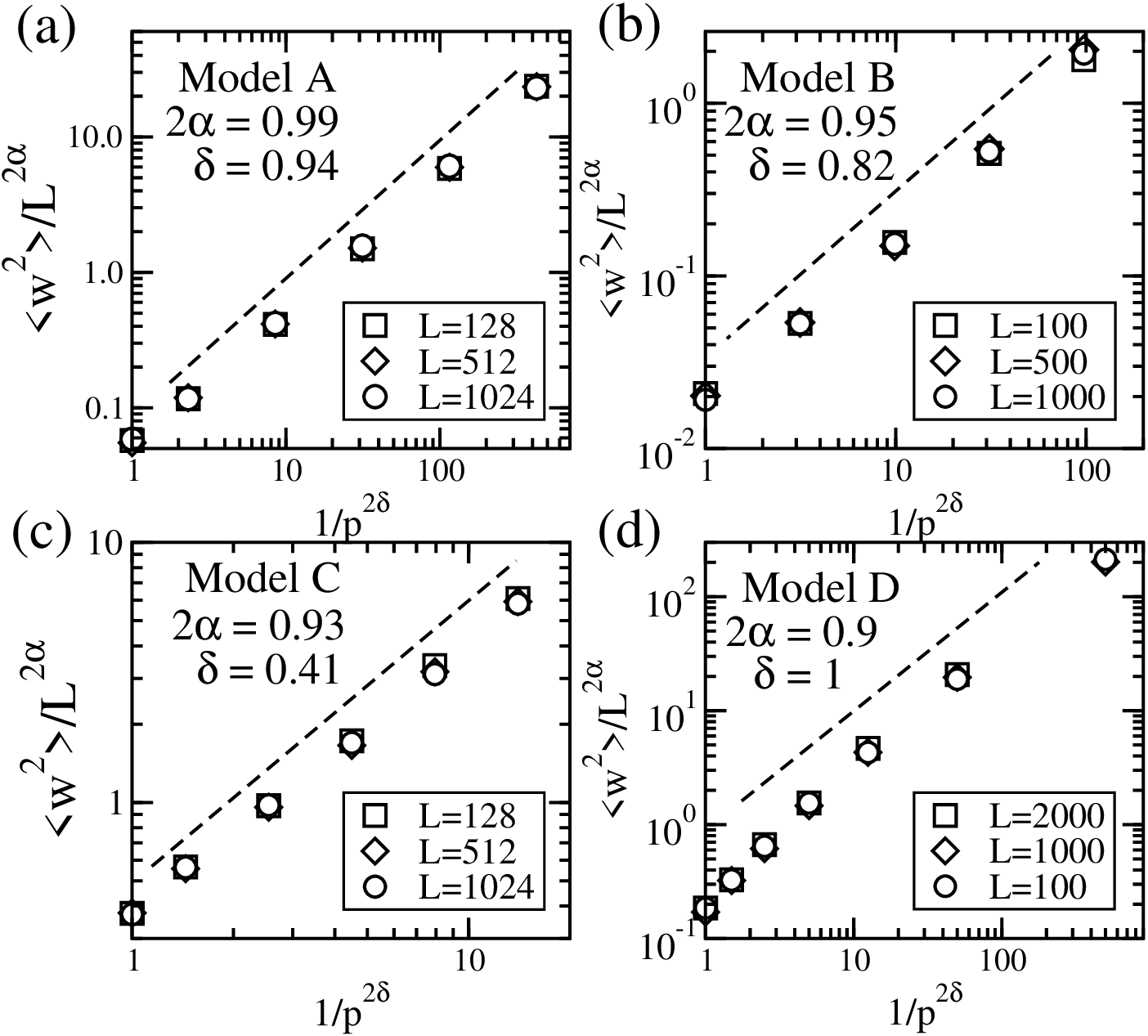}
\caption{\label{fig-2}
Scaled widths at saturation vs parameter
$1/p^{2\delta}$: (a) and (b) are for {\it Models A} and {\it B}, respectively 
(mix of RD with EW processes); 
(c) and (d) are for {\it Models C} and {\it D}, respectively 
(mix of RD with KPZ). Reference lines have slope $1$.
Data are scaled with the exponent values shown here.
}
\end{figure}
%%%%%%%%%%%%%%%%%%%%

In all models, evolutions $w^2(t)$ form two-parameter 
families of curves ($L$ and $p$ being parameters) that for any $p \in (0;1]$ 
look like those in Fig.~1 but with $t_0 \equiv t_0(p) \ge t_0(1)$, 
$t_\times \equiv t_\times (p) \ge t_\times (1)$, and at saturation 
$w^2(p) \ge  w^2(1)$. The curves 
saturate due to {\it only one} component, the pure-correlated deposition, 
since the other component, RD, introduces no correlations. At saturation, 
the observed lateral correlation length is $\xi_\parallel (p) \sim L$ 
and $t_\times (p) \sim L^z$; thus, $\xi_\parallel (p) \sim t_\times^{1/z}(p)$; 
and, the widths scale in $L$ as $w^2(p) \sim L^{2\alpha}$. 
Plots of the scaled widths $\langle w^2(p) \rangle /L^{2\alpha}$  
(Fig.~2) show that they generally scale in $p$ as 
$w^2 \sim L^{2\alpha}/p^{2\delta}$, where $\delta$ is some 
parameter. Is $\delta$ a universal exponent?  
{\it Models A} and {\it B} (Figs.~2a-b) 
may suggest a universal value $\delta =1$ for the RD-EW mix. 
But {\it Models C} and {\it D} (Figs.~2c-d) show 
that $\delta_D \approx 2 \delta_C$.  
Accordingly, $\delta$ is not universal because for the RD-KPZ mix its value 
is clearly related to the technicalities of these models.  
In the RD-EW case there is no reason to believe that $\delta=1$ 
is not accidental. Scale invariance of the EW equation is not sufficient to 
furnish $\delta$.

Since $t_\times (p) \sim L^z$ and 
$w^2 \sim L^{2\alpha}/p^{2\delta}$ for any $p \ne 0$, 
the roughness must scale 
as $w^2 (t)/w^2 \sim F(t(p)/L^z)$. This scaling in 
$L$ collapses all curves $w^2(t)$ to one-parameter families 
($p$ being the only parameter now) presented in Figs.~3a, 4a, 5a and 6a.  
As RD components do not build correlations, this collapse is obtained with 
the scaling laws 
from the corresponding universality classes of processes that build up correlations. 
Explicitly, $z=2\alpha +1$ and $z=2-\alpha$ for blending RD with EW and 
KPZ processes, respectively. To further collapse the data in  
$p$, i.e., to find $t(p)$ in the argument of function $F$, we analyze the invariance 
of the corresponding continuum equations under simultaneous affine transformations:
\begin{equation}
\label{transform-2}
x \to Lx \, , \; h \to h L^\alpha /g \, , \; t \to t L^z / f \, ,
\end{equation}
assuming $g$ and $f$ being arbitrary suitable functions of $p$.

%%%%%%%%  figure 3 %%%%%%%
\begin{figure}[tp]
\includegraphics[width=6.2cm]{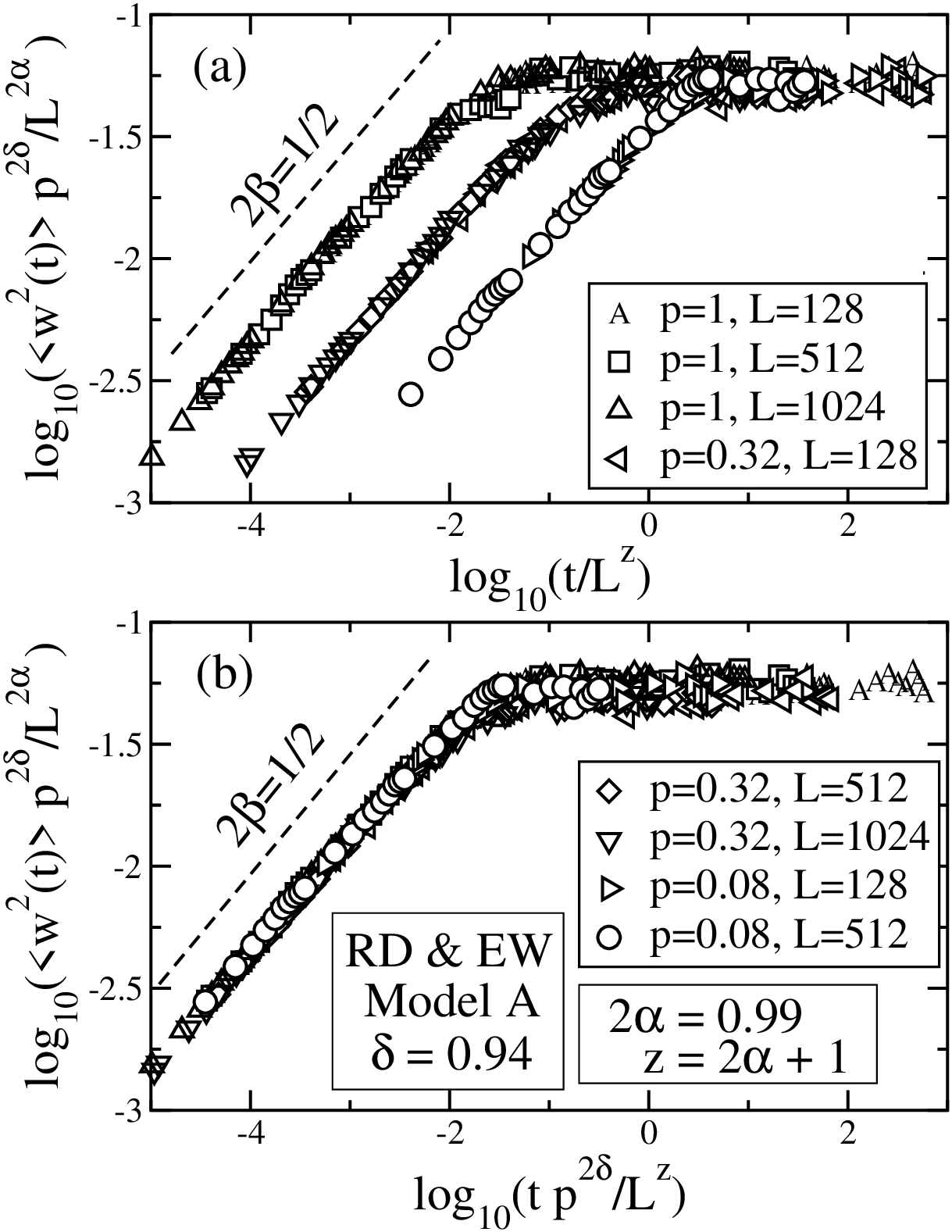}
\caption{\label{fig-3}
Scaling for {\it Model A}: (a) in $L$;
(b) in $p$ of the data in Fig.~(a).
$\alpha$ and $\beta$ are consistent with the EW universality class.
Data labels are common for both figures. $N \approx 100$.
}
\end{figure}
%%%%%%%%%%%%%%%%%%%%

Scaling (\ref{transform-2}) is the superposition of scaling 
(\ref{transform-1}) with
\begin{equation}
\label{transform-3}
x \to x'=x \, , \; h \to h'= h/g(p)\, ,  \; t \to t'= t/f(p) \, .
\end{equation}
Invariance analysis under the component scaling (\ref{transform-1}) 
leads to Eq.~(\ref{FV-1}) and signature-scaling laws of KPZ and EW 
processes. This justifies the data collapse in $L$. 
The component scaling (\ref{transform-3}) transforms Eq.~(\ref{KPZ-1}) to: 
$h'_{t'}=v' + \nu'(p) h'_{x'x'} + (\lambda' (p) /2) h'^2_{x'}+ \eta'(x',t')$, 
where $h'_{t'}=h_t f/g$, $v'=v f/g$, $h'_{x'x'}= h_{xx}/g$, and 
$\eta' (x',t')= \sqrt{f} \eta (x,t)$. Its invariance under  
(\ref{transform-3}) implies:
\begin{eqnarray}
f(p) &=& g^2(p) \label{result-a} \\
\nu'(p) &=& \nu_0 f(p)   \label{result-b} \\
\lambda'(p) &=& \lambda_0 g(p) f(p)  \label{result-c}.
\end{eqnarray}
From scaling at saturation we obtained $g(p)=p^\delta$. Thus,  
the continuum equation for the RD-KPZ mix is
\begin{equation}
\label{KPZ-2}
h_t = v(t) + \nu_0 p^{2\delta} h_{xx}+ 
(\lambda_0/2) p^{3\delta} h^2_x +\eta (x,t).
\end{equation}
In the limits $p \to 1$ and $p \to 0$ Eq.~(\ref{KPZ-2}) describes 
the dynamics of pure processes, i.e., the KPZ-type and RD, respectively. 
Similarly, the invariance of the EW equation 
under scaling (\ref{transform-3}) gives Eqs.~(\ref{result-a})-(\ref{result-b}).
This leads to the continuum equation for the RD-EW mix:
\begin{equation}
\label{EW-2}
h_t = v(t) + \nu_0 p^{2\delta} h_{xx}+ \eta (x,t).
\end{equation}

%%%%%%%%  figure 4 %%%%%%%
\begin{figure}[tp]
\includegraphics[width=6.2cm]{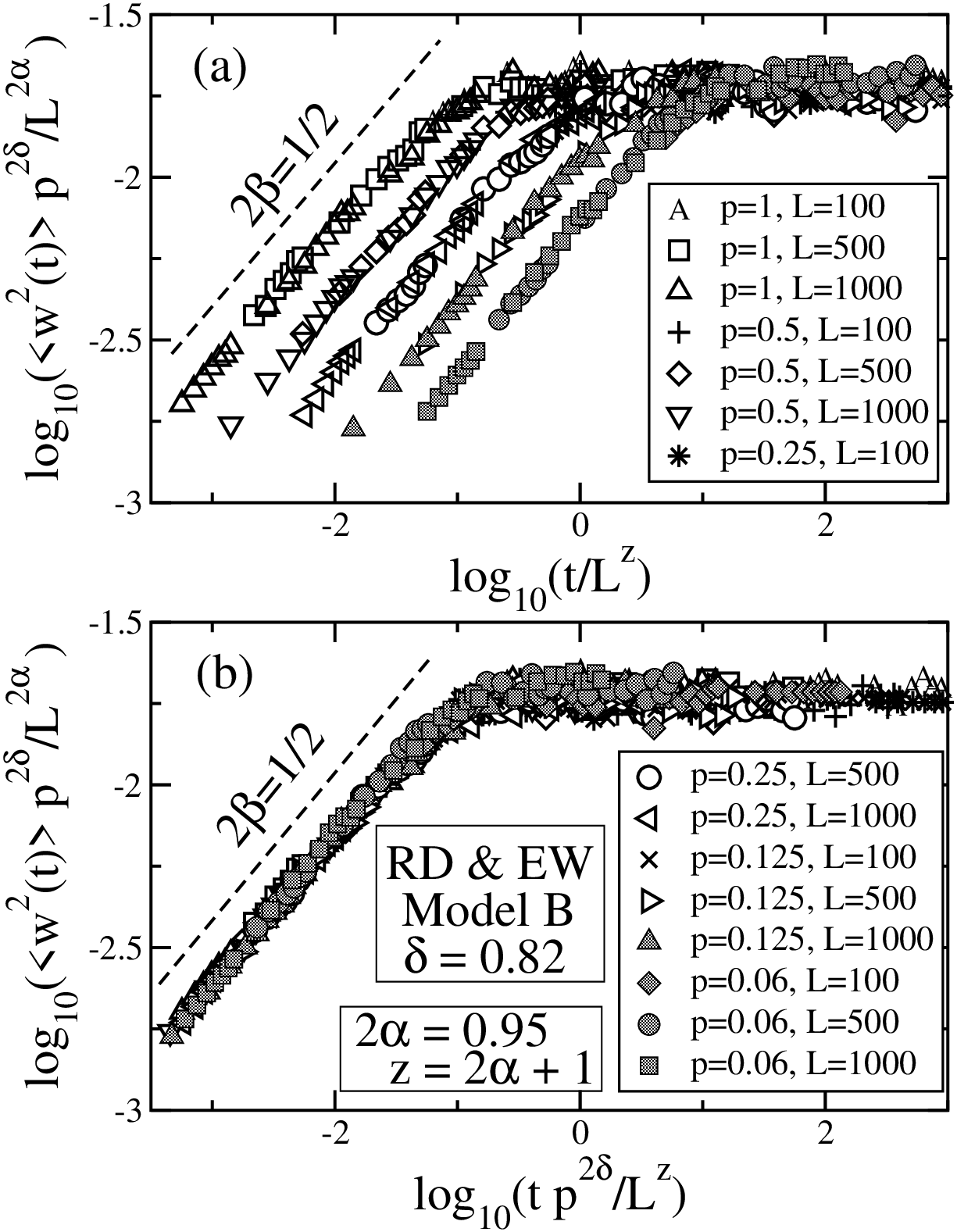}
\caption{\label{fig-4}
Scaling for {\it Model B}: (a) in $L$;
(b) in $p$ of the data in Fig.~(a).
As in {\it Model A}, $\alpha$ and $\beta$ indicate the EW universality class. 
Data labels are common for both figures. $N \approx 100$.
}
\end{figure}
%%%%%%%%%%%%%%%%%%%%

The inverse of the scaling (\ref{transform-2}) is the desired contraction 
that gives the full data collapse described by the FV function. 
The inverse of the scaling (\ref{transform-3}) alone 
($x \to x$, $h \to p^\delta h$, $t \to p^{2\delta}t$) gives $t(p)$ in the argument 
of $F(y)$: $y=p^{2\delta}t/L^z$. 
Finally, the FV scaling for any two-component process, one of which is RD, is
\begin{equation}
\label{FV-2}
w^2(t) = \frac{L^{2\alpha}}{p^{2\delta}} F \left( \frac{p^{2\delta}}{L^z} t \right),
\end{equation}
where $\alpha$ and $z$ are universal exponents of the component process that builds 
up correlations, and $\delta$ is nonuniversal. This result is 
illustrated in Figs.~3b, 4b, 5b and 6b.

Our results, Eqs.~(\ref{KPZ-2})-(\ref{FV-2}), show 
that mixing RD with a correlated growth preserves the universality 
of the correlated growth. Physical justification is in  
the uncorrelated nature of RD. 
As can be seen from Eqs.~(\ref{KPZ-2})-(\ref{EW-2}), RD blending 
reduces the values of coefficients $\nu$ and $\lambda$ relative to 
the original noise strength. In other words, the net outcome is 
a noisier dynamics. The analysis presented here by the examples of 
EW and KPZ processes in ($1+1$) dimensions is easily extended 
to other growth processes in ($1+n$) dimensions. It is enough to notice 
that Eq.~(\ref{result-a}) is generally valid when scaling (\ref{transform-3}) 
applies to growth equations of the type 
$h_t (\vec{x}, t) = (\textrm{operator})h + \eta(\vec{x}, t)$, where 
$\vec{x}$ is $n$ dimensional. 
Hence the conclusion: 
if a correlated growth belongs to a given universality 
class, its mix with RD will remain in the same class. The only effects of the RD 
admixture are the simultaneous dilatations of the fundamental time and height 
scales in accordance with scaling (\ref{transform-3}) (and Eq.~(\ref{result-a})). 
The net consequence of 
these is a slowdown in the dynamics of buildingup the correlations, reflected 
in the change of the lateral correlation length 
$\xi_\parallel (p) \sim t(p)^{1/z} = \xi_\parallel (1)/ \sqrt[z]{f(p)}$. 
In a sense, RD blending is like applying a magnifying glass to the evolution 
curves $w(t)$: the smaller the $p$ the better the magnification. In particular, 
in a two-component growth that mixes RD with either EW or KPZ processes, 
these dilatations explicitly are $h \to h/p^\delta$ and $t \to t/p^{2\delta}$, 
where $\delta$ is nonuniversal and reflects the particulars of the deposition. 
The stretching in time causes the initial nonscaling regime $t_0 (1)$ 
in curves $w(t)$ to be amplified as $t_0 (p) = t_0 (1) /p^{2\delta}$. 
One consequence of this amplification is a clear observation of the RD growth 
(with $\beta = 1/2$) for initial times $t<t_0$ when the growth starts from 
a flat substrate (e.g., observed in \cite{HA03,HMA01,HA01}). 
Note, if $p \to 0$ this initial phase becomes infinitely long as 
this is the limit of RD growth. In simulations, when 
$p$ is known, by a prudent design of a model, magnifying effects of RD blending 
may prove advantageous in revealing hidden features of a correlated growth. 
However, in the laboratory, the presence of randomness in the growth process 
will obscure a clear-cut observation of the expected scaling.

%%%%%%%%  figure 5 %%%%%%%
\begin{figure}[tp]
\includegraphics[width=6.2cm]{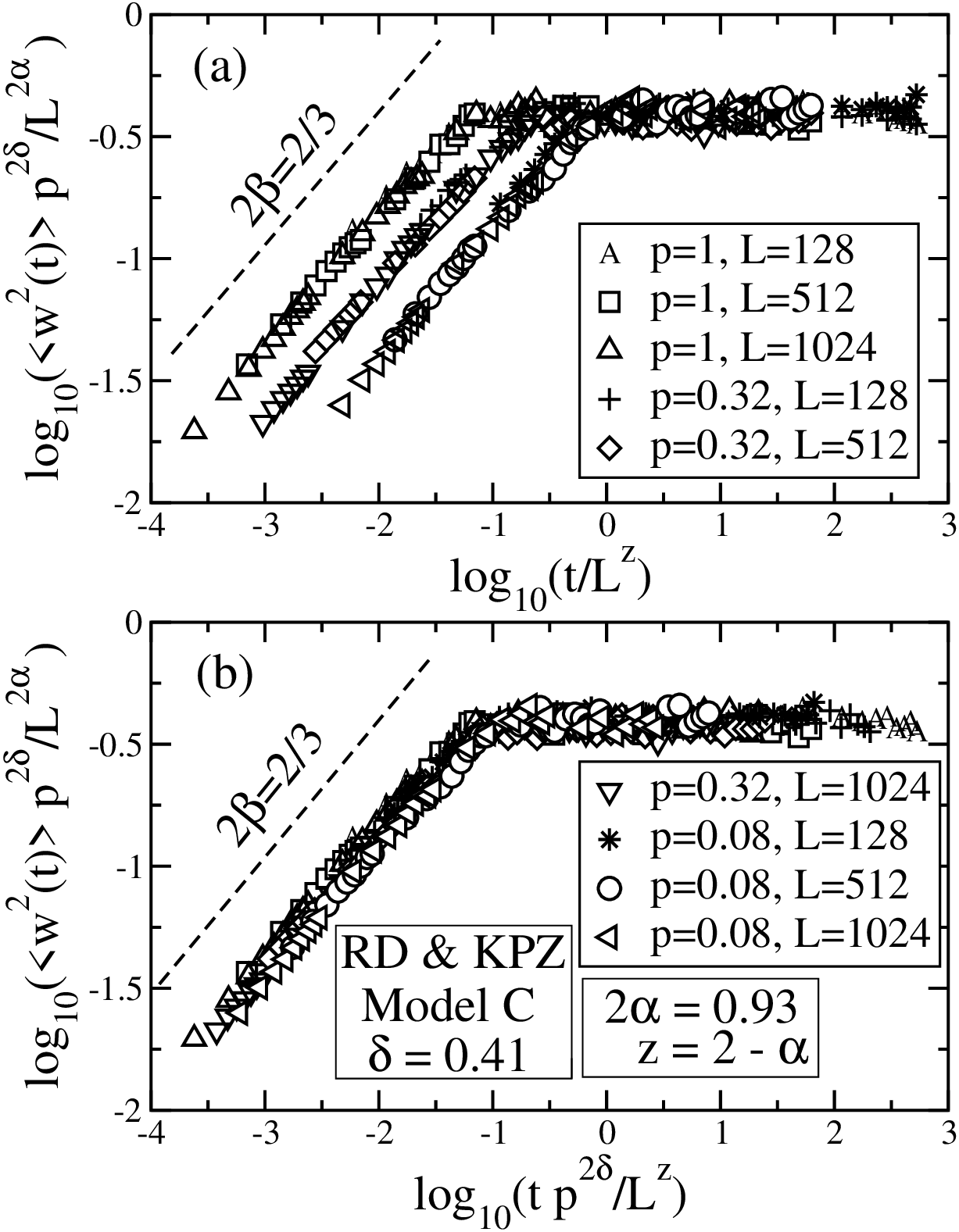}
\caption{\label{fig-5}
Scaling for {\it Model C}: (a) in $L$;
(b) in $p$ of the data in Fig.~(a).
$\alpha$ and $\beta$ are consistent with the KPZ universality.
Data labels are common for both figures.
Here, $N \approx 100$.
}
\end{figure}
%%%%%%%%%%%%%%%%%%%%

%%%%%%%%  figure 6 %%%%%%%
\begin{figure}[!tp]
\includegraphics[width=6.2cm]{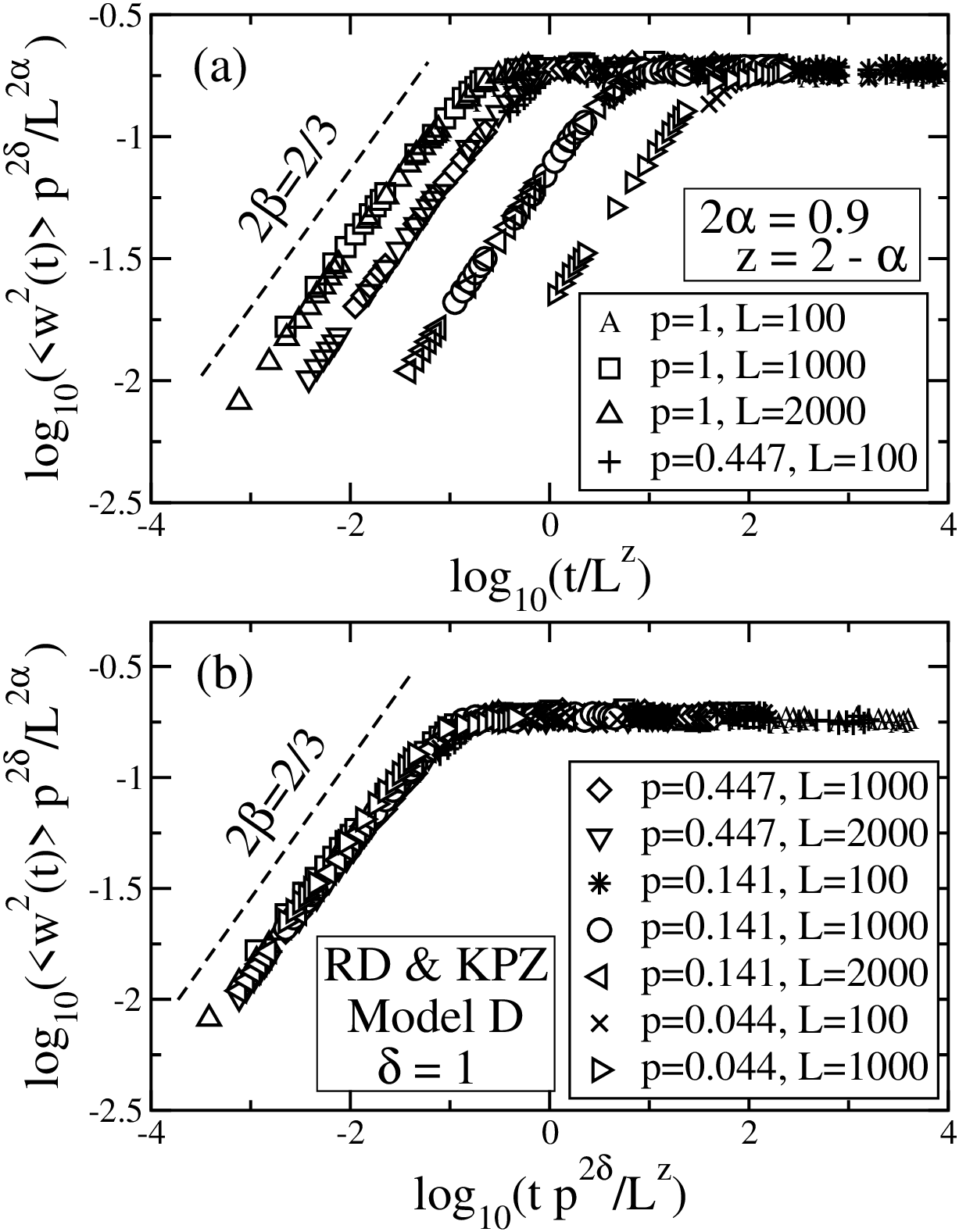}
\caption{\label{fig-6}
Scaling for {\it Model D}: (a) in $L$;
(b) in $p$ of the data in Fig.~(a).
$\alpha$ and $\beta$ indicate the KPZ universality,
but $\delta_D \approx \delta_C/2$.
Data labels are common for both figures.
$N \approx 1000$.
}
\end{figure}
%%%%%%%%%%%%%%%%%%%%

\begin{acknowledgments}
This work is supported by NSF grant DMR-0426488, and by 
the ERC CCS at MSU. It used resources 
of the NERSC Center, supported 
by the Office of Science of the US DoE under contract 
No. DE-AC03-76SF00098.
\end{acknowledgments}

\end{document}